\documentclass[10pt, conference, letterpaper]{IEEEtran}
\IEEEoverridecommandlockouts
\usepackage{cite}
\usepackage{amsmath,amssymb,amsfonts}
\usepackage{algorithmic}
\usepackage{graphicx}
\usepackage{textcomp}
\def\BibTeX{{\rm B\kern-.05em{\sc i\kern-.025em b}\kern-.08em
		T\kern-.1667em\lower.7ex\hbox{E}\kern-.125emX}}
\usepackage{enumerate}
\usepackage{verbatim}
\usepackage[font=small,format=plain,labelsep=period,justification=RaggedRight,singlelinecheck=0]{caption}
\usepackage{cleveref}
\usepackage{hhline}
\usepackage{subfigure}
\usepackage[table]{xcolor}
\usepackage{wrapfig}
\usepackage{multirow}

\begin{document}

	\title{Towards Efficient BLE Mesh: Design of an Autonomous Network Joining Algorithm
	}
	
\author{\IEEEauthorblockN{Clara Nieto-Taladriz, Yuri Murillo  and Sofie Pollin}
 \IEEEauthorblockA{ Department of Electrical Engineering, University of Leuven, Leuven, 3001, Belgium \\
 clara.nietotaladrizmoreno@student.kuleuven.be, yuri.murillo@esat.kuleuven.be}}

	\maketitle
	
	\begin{abstract}     
    
     \par The Internet of Things (IoT) opens the doors to a digital revolution, but requires a robust protocol to wirelessly interconnect a large number of devices. Bluetooth Low Energy (BLE) Mesh emerges as a suitable candidate, solving the range limitations of original BLE. Nevertheless, the main limitation is shifted to the scatternet architecture: how the considerable number of end systems are interconnected to ensure the network efficiency and scalability. As of today, some timid solutions have emerged, though the most relevant parameters for the best parent selection in BLE mesh network joining procedures have not been identified yet. In this work, we perform a thorough exploration of different device and environmental parameters in a BLE mesh testbed to analyze their impact in the overall network figures of merit, such as end-to-end delay and packet delivery ratio (PDR). According to the inferred relevance, the second part of the work implements and measures the proposed parent selection algorithm. The implementation is based on the open source Fruitymesh protocol and is used as a baseline. The results reflect an enhancement in the network scalability and fairness, accomplishing a delay and PDR improvement of 26\% and 10\% respectively, and the avoidance of saturated branches of 24\%.

	\end{abstract}
	
	
	\section{Introduction } \label{Sec1}
    \par The integration of wireless networks in our society acquires promising forecasts, powered by the IoT emergence \cite{Forbes}. However, they require the interconnection of a massive number of devices. Bluetooth Low Energy (BLE) emerges forcefully for this purpose \cite{Collota2018} thanks to its low cost and strong presence in the marketplace. Nonetheless, BLE is a single-hop technology with a coverage range of scarcely 10-15 meters in indoor applications \cite{BLEcoverage}. The BLE mesh \cite{BLEMesh} release granted BLE with a mesh structure and multi-hop communication, so the coverage range can be easily augmented by enlarging the scatternet.
    
    \par  The BLE mesh relaying is flooding-based. Albeit is simple and straightforward, it suffers from high packet collision, unnecessary high power consumption and its set of rules does not allow fine-tuning. As a consequence, the drawback is shifted to the scatternet structure and how the different end systems are connected within the network to ensure the network efficiency. 
    
    \par Shifting BLE mesh towards a routing-based, connected network may solve this issue. Nonetheless, BLE nodes do not all share the same effectiveness in relaying messages nor holding new nodes, due to the heterogeneity germane to IoT applications. Therefore, the different routing-based alternatives also reinforce the relevance of the topology for the overall network performance.

  \par A multi-hop protocol and the handling of messages with different priorities were already implemented in our BLE mesh testbed \cite{MurilloMangeYuri2018AMLA}. Hereby, this work proposes an algorithm to join the BLE network efficiently. Our contribution is twofold. First a wider and thoroughgoing analysis was performed to identify the most relevant parameters in the network joining process. Thereupon, they were weighted according to the inferred relevance and the resulting algorithm was measured and compared with the current procedure.

    \section{Current Framework} \label{Sec3}
    
\par
\begin{figure}[htbp]
\includegraphics[width=61mm]{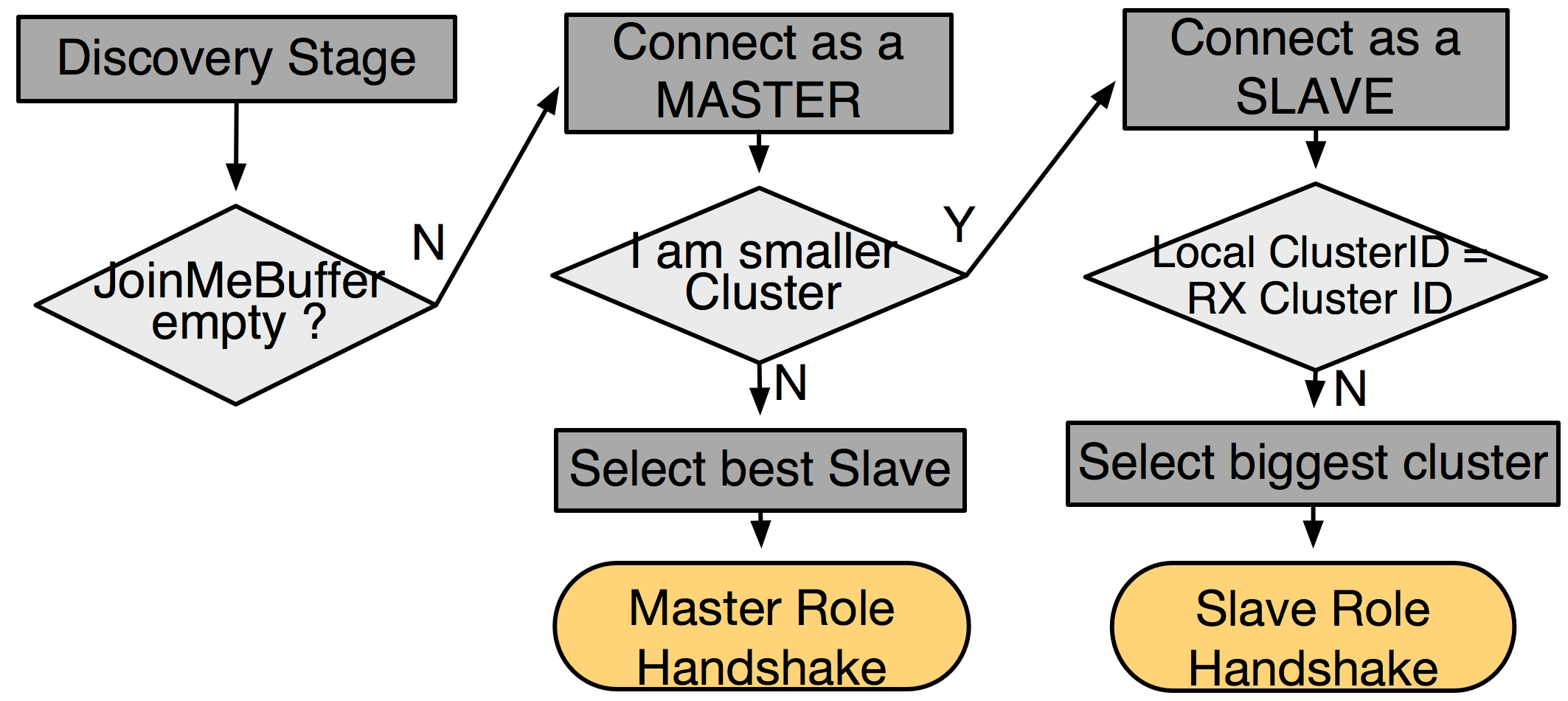}
\centering
\caption{Simplified flowchart of the FruityMesh network build up.}
\label{buildUp}
\end{figure}

 Currently, the BLE nodes use simultaneously two mesh protocols \cite{MurilloMangeYuri2018AMLA}: Trickle \cite{Trickle}, a flooding scheme used for broadcasting advertisements; and FruityMesh \cite{FruityMesh}, a connection-oriented protocol used for regular data to reduce the energy consumption. FruityMesh is a clustering protocol that tends to convergence into a large and common network, which emphasizes the relevance that the location of each end system acquires to guarantee the network performance. However, this issue is not considered with the current procedure (Fig. \ref{buildUp}).
    
\section{Proposed BLE mesh network joining framework} \label{Sec4}
    
     \par
\begin{figure}[htbp]
\includegraphics[width=71mm]{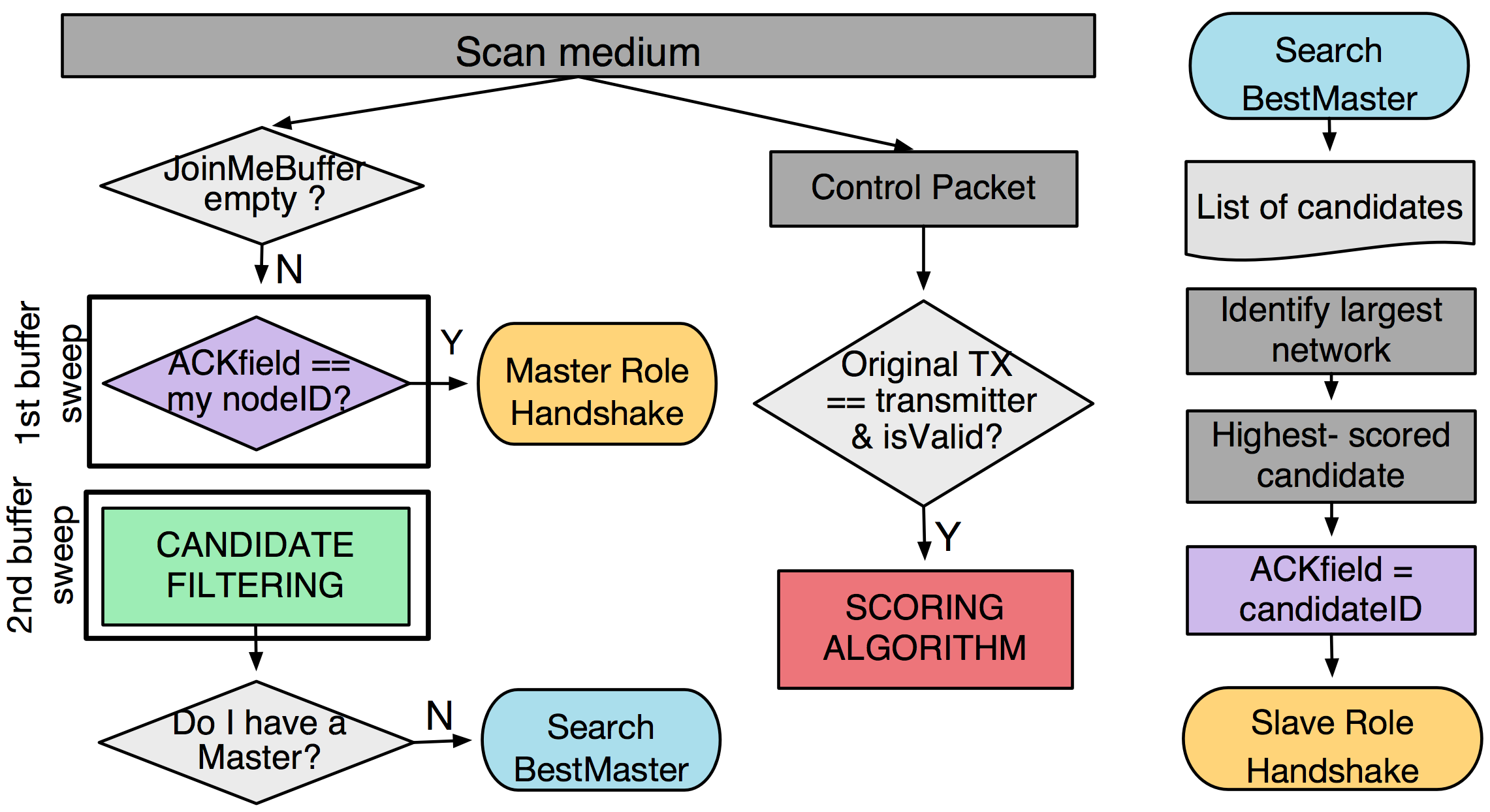}
\centering
\caption{Simplified flowchart of the proposed network build up.}
\label{buildUpPRO}
\end{figure}

 The proposed algorithm (Fig. \ref{buildUpPRO}) filters the neighbors suitable for a connection using the information extracted from broadcasted status messages and scores them. Afterwards, the highest-scored within the biggest cluster is chosen as a parent, and the $ACKfield$ of the joinMe packet acquires its identifier. This way, the node is able to explicitly specify its preferable connection.

\par
\begin{figure}[htbp]
\centering
\fbox{\includegraphics[width=56mm]{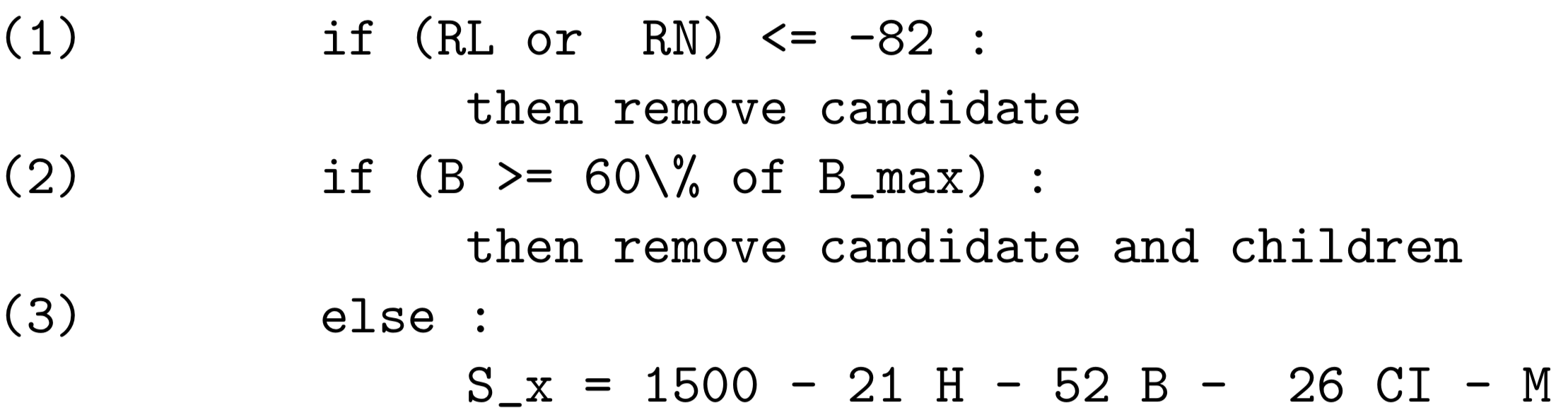}}
\caption{Pseudocode of the proposed scoring algorithm.}
\label{fig_codeScore}
\vspace*{-0.5cm}
\end{figure}

 The proposed scoring algorithm (Fig. \ref{fig_codeScore}) is compound by the slaves number (M), hops to the sink (H), buffer occupancy (B), connection interval (CI), link received signal strength (RSSI) (RL) and the RSSI between the neighbor and its master (RN).

The numerical values were chosen after a thorough analysis, measuring empirically the impact and relevance of each scoring parameter in the BLE mesh testbed performance. The latter is evaluated through two main figures of merit: the end-to-end delay and PDR. The chosen weights work as a ballpark.

\section{ Experimental results} \label{Sec4}

\begin{figure}[h]
\centering
\subfigure[\label{fig_delay}]{\includegraphics[width=30mm]{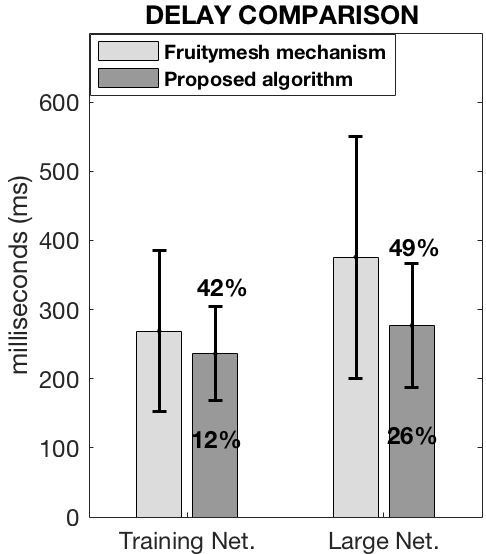}}
 \subfigure[\label{fig_PDR}]{\includegraphics[width=27.5mm]{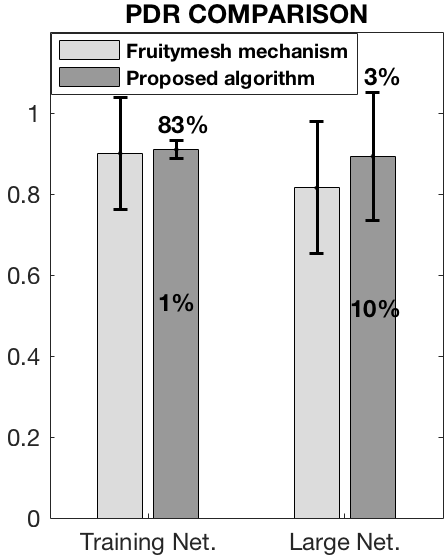}}
 \subfigure[\label{fig_sat}]{\includegraphics[width=29mm]{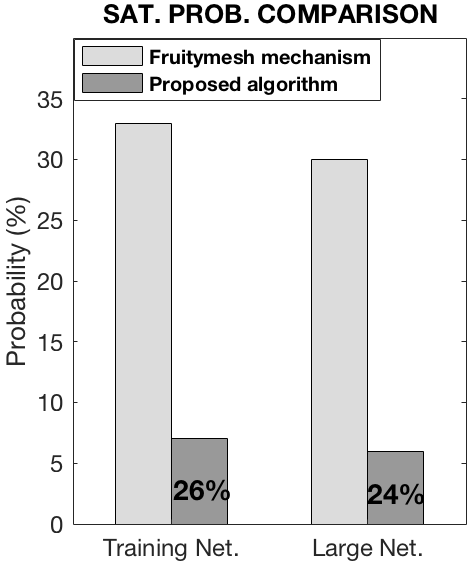}}
\caption{Joining procedures comparison.}
\label{fig_comp}
\vspace{-0.4cm}
\end{figure}

The proposal was deployed in the training (11 BLE nodes) and in larger random-generated networks (16 nodes), using the original FruityMesh procedure as a baseline. First columns of Table \ref{table_final} show the delay and PDR mean ($\mu$) and deviation ($\sigma$) empirically measured in the new node. $\% Sat$ is the probability of the new device working in saturated branches, $\overline{N_{Hops}}$ the mean distance occupied by the latter and $avoid\_Sat$ denotes the avoided congested branches. The comparison in terms of delay, PDR and probability of saturation is depicted in Fig. \ref{fig_comp}.


\begin{table}[h]
\centering
\resizebox{8.4cm}{!}{
\begin{tabular}{cccccccc}
\hline
\rowcolor[HTML]{C1E1E8} 
\multicolumn{1}{|c|}{\cellcolor[HTML]{C1E1E8}} & \multicolumn{1}{c|}{\cellcolor[HTML]{C1E1E8} $\mu_d$} & \multicolumn{1}{c|}{\cellcolor[HTML]{C1E1E8}$\sigma_{d}$} & \multicolumn{1}{c|}{\cellcolor[HTML]{C1E1E8}$\mu_{PDR}$} & \multicolumn{1}{c|}{\cellcolor[HTML]{C1E1E8}$\sigma_{PDR}$} & \multicolumn{1}{c|}{\cellcolor[HTML]{C1E1E8}\% Sat } & \multicolumn{1}{c|}{\cellcolor[HTML]{C1E1E8}avoid\_Sat} & \multicolumn{1}{c|}{\cellcolor[HTML]{C1E1E8}$\overline{N_{Hops}}$} \\ \hline

\rowcolor[HTML]{EFEFEF} 
\multicolumn{1}{|c|}{\cellcolor[HTML]{C0C0C0}} & \multicolumn{7}{c|}{\cellcolor[HTML]{EFEFEF}Applying the \textbf{Proposed} algorithm} \\ \cline{2-8} 
\multicolumn{1}{|c|}{\cellcolor[HTML]{C0C0C0}} & \multicolumn{1}{c|}{236} & \multicolumn{1}{c|}{68} & \multicolumn{1}{c|}{0.91} & \multicolumn{1}{c|}{0.02} & \multicolumn{1}{c|}{7\%} & \multicolumn{1}{c|}{71\%} & \multicolumn{1}{c|}{2.4} \\ \cline{2-8}

\rowcolor[HTML]{EFEFEF} 
\multicolumn{1}{|c|}{\cellcolor[HTML]{C0C0C0}} & \multicolumn{7}{c|}{\cellcolor[HTML]{EFEFEF}Applying the \textbf{Fruitymesh} mechanism} \\ \cline{2-8} 
\multicolumn{1}{|c|}{\multirow{-4}{*}{\cellcolor[HTML]{C0C0C0}\begin{tabular}[c]{@{}c@{}}TRAINING\\ NETWORK\end{tabular}}} & \multicolumn{1}{c|}{269} & \multicolumn{1}{c|}{117} & \multicolumn{1}{c|}{0.90} & \multicolumn{1}{c|}{0.14} & \multicolumn{1}{c|}{33\%} & \multicolumn{1}{c|}{-} & \multicolumn{1}{c|}{2.6} \\ \hline

\rowcolor[HTML]{EFEFEF} 
\multicolumn{1}{|c|}{\cellcolor[HTML]{C0C0C0}} & \multicolumn{7}{c|}{\cellcolor[HTML]{EFEFEF}Applying the \textbf{Proposed} algorithm} \\ \cline{2-8} 
\multicolumn{1}{|c|}{\cellcolor[HTML]{C0C0C0}} & \multicolumn{1}{c|}{277} & \multicolumn{1}{c|}{90} & \multicolumn{1}{c|}{0.89} & \multicolumn{1}{c|}{0.158} & \multicolumn{1}{c|}{6\%} & \multicolumn{1}{c|}{70\%} & \multicolumn{1}{c|}{2.7} \\ \cline{2-8}
\rowcolor[HTML]{EFEFEF}

\multicolumn{1}{|c|}{\cellcolor[HTML]{C0C0C0}} & \multicolumn{7}{c|}{\cellcolor[HTML]{EFEFEF}Applying the \textbf{Fruitymesh} mechanism} \\ \cline{2-8} 
\multicolumn{1}{|c|}{\multirow{-4}{*}{\cellcolor[HTML]{C0C0C0}\begin{tabular}[c]{@{}c@{}}LARGER\\  NETWORKS\end{tabular}}} & \multicolumn{1}{c|}{376} & \multicolumn{1}{c|}{175} & \multicolumn{1}{c|}{0.82} & \multicolumn{1}{c|}{0.163} & \multicolumn{1}{c|}{30\%} & \multicolumn{1}{c|}{-} & \multicolumn{1}{c|}{3.5} \\ \hline
\end{tabular}
}
\caption{Fruitymesh and proposed algorithm comparison.}
\label{table_final}
\end{table}

\par In the training network both algorithms select the same $\overline{N_{Hops}}$. Since our algorithm considers the CI and chooses chiefly empty buffers, it attains a delay enhancement of 12\% (Fig. \ref{fig_delay}), which conspicuously improves in larger networks, reaching the 26\%. The significant reduction of $\sigma_d$ evinces a higher precision in choosing parents that guarantee low delays. 


\par The PDR improvement accomplishes the 10\% in larger networks (Fig. \ref{fig_PDR}) since the CI consideration allows selecting nodes with buffers that empty faster. Furthermore, the noteworthy $\sigma_{PDR}$ reduction in small networks also denotes the tendency of choosing the best parent. 

\par The proposed algorithm wisely discards bottlenecks. Moreover, discarding nodes without an empty buffer but selecting a child introduces the fairness concept. Notwithstanding, if at the decision instant a device has a buffer occupancy fairly close to the threshold, choosing a child will fail on the saturation avoidance. This drawback justifies the meager $\sigma_{PDR}$ reduction in larger scenarios. Nevertheless, it corresponds to an infrequent case. In Table \ref{table_final} $avoid\_Sat$ indicates that the 70\% of branches with high traffic load are successfully avoided.

\par Fig. \ref{fig_sat} represents a reduction of 25\% in the probability of operating in saturated environments. Since a link suffering from congestion flagrantly affects the whole branch, the improvement observed in the last two graphics in the new node can be extrapolated as an enhancement of the entire network.
 
\section{Conclusions} \label{Sec6}

\par This paper proposes a network building scheme for BLE mesh networks. Albeit in general enlarging the scatternet will always worsen the global performance, a good master and network position allows mitigating the adverse effects. The proposed algorithm allows integrating a new device under the principles of fairness and scalability. In the vast majority of cases, the new node joins the scatternet without jeopardizing the overall network performance and being capable of performing a befitting operation in terms of delay and PDR.

\par The scoring algorithm values work as a ballpark. Albeit they may differ from the optimum, they are  accurate enough to improve the overall network performance figures of merit.

	\addcontentsline{toc}{section}{References}


\begin{thebibliography}{1}
\providecommand{\url}[1]{#1}
\csname url@samestyle\endcsname
\providecommand{\newblock}{\relax}
\providecommand{\bibinfo}[2]{#2}
\providecommand{\BIBentrySTDinterwordspacing}{\spaceskip=0pt\relax}
\providecommand{\BIBentryALTinterwordstretchfactor}{4}
\providecommand{\BIBentryALTinterwordspacing}{\spaceskip=\fontdimen2\font plus
\BIBentryALTinterwordstretchfactor\fontdimen3\font minus
  \fontdimen4\font\relax}
\providecommand{\BIBforeignlanguage}[2]{{%
\expandafter\ifx\csname l@#1\endcsname\relax
\typeout{** WARNING: IEEEtran.bst: No hyphenation pattern has been}%
\typeout{** loaded for the language `#1'. Using the pattern for}%
\typeout{** the default language instead.}%
\else
\language=\csname l@#1\endcsname
\fi
#2}}
\providecommand{\BIBdecl}{\relax}
\BIBdecl

\bibitem{Forbes}
\BIBentryALTinterwordspacing
Forbes, ``{2017 Roundup Of Internet Of Things Forecasts}.'' [Online].
  Available:
  \url{https://www.forbes.com/sites/louiscolumbus/2017/12/10/2017-roundup-of-internet-of-things-forecasts}
\BIBentrySTDinterwordspacing

\bibitem{Collota2018}
M.~Collotta, G.~Pau, T.~Talty, and O.~K. Tonguz, ``Bluetooth 5: A concrete step
  forward toward the {IoT},'' \emph{IEEE Communications Magazine}, vol.~56,
  no.~7, pp. 125--131, JULY 2018.

\bibitem{BLEcoverage}
P.~D. Marco, R.~Chirikov, P.~Amin, and F.~Militano, ``{Coverage analysis of
  Bluetooth low energy and IEEE 802.11ah for office scenario},'' in \emph{2015
  IEEE 26th Annual International Symposium on Personal, Indoor, and Mobile
  Radio Communications (PIMRC)}, Aug 2015, pp. 2283--2287.

\bibitem{BLEMesh}
\BIBentryALTinterwordspacing
{Bluetooth SIG}, ``{BLE mesh profile specification 1.0},'' July 2017. [Online].
  Available: \url{https://www.bluetooth.com/specifications/mesh-specifications}
\BIBentrySTDinterwordspacing

\bibitem{MurilloMangeYuri2018AMLA}
Y.~Murillo~Mange, B.~Reynders, A.~Chiumento, and S.~Pollin, ``{A Multiprotocol
  Low-Cost Automated Testbed for {BLE} Mesh},'' \emph{IEEE Communications
  Magazine}, 2018.

\bibitem{Trickle}
P.~Levis, N.~Patel, D.~Culler, and S.~Shenker, ``{Trickle: A Self-regulating
  Algorithm for Code Propagation and Maintenance in Wireless Sensor
  Networks},'' in \emph{Proceedings of the 1st Conference on Symposium on
  Networked Systems Design and Implementation - Volume 1}, ser. NSDI'04.\hskip
  1em plus 0.5em minus 0.4em\relax Berkeley, CA, USA: USENIX Association, 2004,
  pp. 2--2.

\bibitem{FruityMesh}
\BIBentryALTinterwordspacing
FruityMesh, ``Fruitymesh,'' accessed 22/08/2018. [Online]. Available:
  \url{https://github.com/mwaylabs/fruitymesh/wiki}
\BIBentrySTDinterwordspacing

\end{thebibliography}
\end{document}